\documentclass[prb,preprint]{revtex4-1} 

\usepackage{amsmath}  
\usepackage{amsfonts} 
\usepackage{graphicx} 

\begin{document}


\title{What happens if measure the electron spin twice?}

\author{Y. C. Zou}
\email{zouyc@hust.edu.cn} 
\affiliation{School of Physics, Huazhong University of Science and Technology, Wuhan 430074, China}

\date{\today}

\begin{abstract}
The mainstream textbooks of quantum mechanics explains the quantum state collapses into an eigenstate in the measurement, while other explanations such as hidden variables and multi-universe deny the collapsing. 
Here we propose an ideal thinking experiment on measuring the spin of an electron with 3 steps.  It is simple and straightforward, in short, to measure a spin-up electron in x-axis, and then in z-axis.  Whether there is a collapsing predicts different results of the experiment. The future realistic experiment will show the quantum state collapses or not in the measurement. 
\end{abstract}


\maketitle 

\section{Introduction} 
In most of the textbooks of quantum mechanics \cite{wichmann1974quantum, CohenTannoudji1986, sakurai1995modern, liboff2003introductory, zhao2008, griffiths2016introduction}, authors choose the explanation of the measurement that the quantum state collapses into an eigenstate, which is taken as one of the hypothesises of the quantum mechanics. The probability for collapsing into each eigenstate is given by the absolute square of the inner product between the eigenstate vector and the state vector, which is consistent with all the experiments.
However, the collapsing is still criticized by its `unnatural' property. The alternative explanations were proposed, such as the hidden variables \cite{bohm1952suggesteda, bohm1952suggestedb}, the decoherence theory \cite{Zeh1970} and many worlds \cite{everett1957relative} etc.
The  `unnatural' fact also leads to further thinking and paradoxes, like the Schr\"odinger's cat \cite{schrodinger1935present}, the EPR paradox \cite{einstein1935can},  Bell inequality \cite{Bell1964a} and  the Wheeler’s  delayed-choice experiment \cite{wheeler1984choice} etc. There are many recent experiments or thinking proposals which are trying to figure out the quantum reality, or the naturality of the quantum theory \cite{0143-0807-4-2-008, ou1992realization, pan2000experimental, jacques2007experimental, pusey2012reality, fuchs2014introduction, frauchiger2016single, weinberg2012collapse, bassi2013models}.

Even all the predictions of the collapsing explanation are consistent with the experiments, it does not necessarily indicate that there is no experiment can violate the explanation.
Here we suggest a possible experiment by measuring the spin of an electron, and the predicted result is testable. This experiment may be used to determine whether there is a collapsing.

\section{Processes of the experiment}
In the ideal thinking experiment, we suppose the spin of the electron can be prepared and be measured ideally. The experiment is designed as the following.

Step 1, prepare an electron at state with spin in positive z-axis $|z_+\rangle$ (i.e., spin-up state).

Step 2, measure the spin in x-axis, and record the result as  $|x_+\rangle$ or  $|x_-\rangle$. 

Step 3, measure the spin in z-axis, and take the result ($|z_+\rangle$ or $|z_-\rangle$) for comparison.

Then, repeat the 3 steps for many times, to see the chance of appearing $|z_+\rangle$ and $|z_-\rangle$.

The key point is that in the step 2, whether the observer is in a superposition state of ``recorded $|x_+\rangle$" and ``recorded $|x_-\rangle$", or is in a collapsed state of either ``recorded $|x_+\rangle$" or ``recorded $|x_-\rangle$". If the observer is "us", we may think we are in a collapsed state. But in the viewpoint of an outlier, he/she may think the observer (either a cat, a machine or a human being) is in a superposition state, i.e., the  Schr\"odinger's cat state. Which one is true? A physical experiment should not depend on who is the operator of the experiment.  Fortunately, the two points of view lead to different results.

To clarify, we design two versions of the experiment (actually two different points of view). The results can be used for identification by the performed experiment.

In version A, we engage Alice to do the experiment from step 1 to step 3, and we stand on the point of view of Alice.
 As one can imagine, for each measurement, the result should be 1/2 chance of  $|x_+\rangle$ (or $|z_+\rangle$), and 1/2 chance of $|x_-\rangle$ (or $|z_-\rangle$) in step 2. After step 2, the electron is either in  $|x_+\rangle$ or in  $|x_-\rangle$ with 100\% chance. Whatever it starts from $|x_+\rangle$ or  $|x_-\rangle$, in step 3, we get the electron 1/2 chance in  $|z_+\rangle$, and 1/2 chance in $|z_-\rangle$, which is the result.



Let's express the states and measurements in Pauli spin matrix in the z-axis base\cite{liboff2003introductory}. I.e., 
\begin{equation}
 |z_+\rangle=\left(
  \begin{array}{l} 1 \\ 0 \end{array}
 \right),
 |z_-\rangle=\left(
  \begin{array}{l} 0 \\ 1 \end{array}
 \right),
\end{equation}
are the two spin vectors in z-axis,
\begin{equation}
 |x_+\rangle=\frac{1}{\sqrt{2}}\left(
  \begin{array}{l} 1 \\ 1 \end{array}
 \right),
 |x_-\rangle=\frac{1}{\sqrt{2}}\left(
  \begin{array}{l} 1 \\ -1 \end{array}
 \right),
\end{equation}
are the two spin vectors in x-axis, and
\begin{equation}
 |y_+\rangle=\frac{1}{\sqrt{2}}\left(
  \begin{array}{l} 1 \\ i \end{array}
 \right),
 |y_-\rangle=\frac{1}{\sqrt{2}}\left(
  \begin{array}{l} 1 \\ -i \end{array}
 \right),
\end{equation}
are the two spin vectors in y-axis. `+' donates the direction of the axis, and `-' donates the anti-direction.

The Pauli matrices
\begin{equation}
 \sigma_z=\left(
  \begin{array}{ll} 1 & 0 \\ 0 & -1 \end{array}
 \right),
 \sigma_x=\left(
  \begin{array}{ll} 0 & 1 \\ 1 & 0 \end{array}
 \right),
 \sigma_y=\left(
  \begin{array}{ll} 0 & -i \\ i & 0 \end{array}
 \right),
\end{equation}
act as the measurement operators in z-axis, x-axis and y-axis respectively.

For step 2, it is equivalent to the operation: $\sigma_x |z_+\rangle $, which equals to $ (\frac{1}{\sqrt{2}}|x_+\rangle-\frac{1}{\sqrt{2}}|x_-\rangle)$. Alice's measurement forces the state collapsing into either $|x_+\rangle$ or $|x_-\rangle$ with equal chances \cite{heisenberg1958physics}. Step 3 starts from a state either $|x_+\rangle$ or $|x_-\rangle$, and both will get $|z_+\rangle$ or $|z_-\rangle$ in equal chances as her final report (probability equals to 1/2). This result is explicitly adopted in several textbooks, such as, example 11 on page 54 of reference \cite{zhao2008}, and \S 4.4.1 of reference \cite{griffiths2016introduction}.

In experiment version B, it is the same 3 steps as done in version A. The only difference is the point of view. Now we engage Bob as an outer observer, and take Alice together with the whole experiment A as a black box (we stand on  Bob's point of view).
Step 1, it is the same in Bob's view.
Step 2, it is different. In Bob's view, Alice is now in the superposition state, namely, her result is $\sigma_x |z_+\rangle = (\frac{1}{\sqrt{2}}|x_+\rangle-\frac{1}{\sqrt{2}}|x_-\rangle)$ instead. 
Step 3, measuring the spin of z-axis is equivalently to do the following operation: $\sigma_z (\sigma_x |z_+\rangle) $, which is equal to $ -|z_-\rangle$.
Therefore, Bob gets the result always being  $|z_-\rangle$.
 It is clearly different from experiment  version A, which is waiting for distinguishing by the real experiment.

\section{Discussion}

In these two versions of the experiment, the experiment itself is actually the same, while the only difference is the viewpoint. The future performed experiment will show which viewpoint is correct.
According to historical quantum predictions and realizations, such as EPR paradox \cite{einstein1935can} and its experiment \cite{ou1992realization}, Bell's inequality \cite{Bell1964a} and its experimental violation \cite{clauser1978bell}, GHZ paradox \cite{greenberger1989going} and its experiment \cite{pan2000experimental}, Wheeler's delayed choice experiment \cite{wheeler1984choice,jacques2007experimental}, one may bet more on the Bob's view  correct. If it is true, what is wrong with Alice's view?

Notice the only difference  between viewpoints of Alice and Bob is that: Alice thinks the state collapses into an eigenstate on the x-axis (z-axis), while Bob thinks Alice is still in a superposition state. Therefore, Alice may think wrong, even if she does observe the spin pointing to $|x_+\rangle$, and write ``$|x_+\rangle$" in her record (e.g. in step 2), she is still in a superposition state.
That is to say, in the measurement, the quantum state is still a quantum state, but the observer can only be aware of the eigenstate, and consequently the observer goes into a superposition of aware of different eigenstates. 

With each measurement, the world maybe splits into many worlds, which is consistent with the many worlds explanation. Back to the thinking experiment of measuring the electron's spin, in step 2, the word splits to two worlds. One is with $|x_+\rangle$, and the other is with $|x_-\rangle$. However, these two worlds are connected. The connection appears in the step 3. If there is no connection, the two worlds will continue splitting into 4 worlds in the step 3, with two of them $|z_+\rangle$ and two of them $|z_-\rangle$. However, if the viewpoint of experiment version B is correct, the worlds should combine into one world  only with $|z_-\rangle$.

Schr\"odinger's cat \cite{schrodinger1935present} is an amazing paradox in quantum mechanics. In the common view (or the Copenhagen interpretation)\cite{heisenberg1958physics}, the cat's state cannot be identified, because once it is observed, the cat's superposition state collapses into a normal state. And the cat cannot be replaced by a human as the human is aware of the state collapsing \cite{Wigner1961-WIGROT}. This thinking experiment may identify the existence of the Schr\"odinger's cat state.

In step 2, one can measure in y-axis instead of in x-axis, and all the results are kept the same. This is reasonable, as the x-axis and y-axis is equivalent in the eye of z-axis.
One more interesting thing is that the measurement does effect the state. Otherwise, the measured results should be spin-up again in step 3, while the predicted results are spin-downs, i.e., $|z_-\rangle$.

On the other hand, if the experiment result is half being $|z_+\rangle$ and half being $|z_-\rangle$ as predicted by the version A, this may indicate that the macroscopic Schr\"odinger's cat state does not exist. Notice in the step 2, Alice can be replaced by a machine to automatically perform and record. That is to say, whatever the electron spin is measured by a human being, or a cat, or an instrument, the system of the spin and the measurer maybe collapses into an eigenstate as suggested by the Copenhagen interpretation, or becomes decoherent as suggested by the decoherence theory.

\section{Conclusion}
In this work, a possible experiment to identify whether the quantum state collapses in the measurement is proposed. With  an electron is prepared in spin-up state initially, we can measure the spin of the electron in x-axis and then in z-axis again. Then repeat the steps many times to see the final results being spin-up or spin-down. If each measurement forces the electron collapsing into an eigenstate, we will get the final results half being spin-up and half being spin-down. While if measurement does not force the collapsing and the superposition  Schr\"odinger's cat state is real, we will get the final spin in z-axis always being spin-down. It is  easy to be distinguished in the future performed experiment. Though this kind of experiment is straightforward and seems easy to perform, no one has ever really done it. Therefore, a real experiment is highly expected.

\end{document}